# Resistivity and Thermal Conductivity of an Organic Insulator β'–EtMe$_3$Sb[Pd(dmit)$_2$]$_2$

**Authors:**

Minoru Yamashita[1,*], Yuki Sato[2†], Yuichi Kasahara[2], Shigeru Kasahara[2‡], Takasada Shibauchi[3] and Yuji Matsuda[2]

[1] The Institute for Solid State Physics, University of Tokyo, Chiba 277-8581, Japan
[2] Department of Physics, Kyoto University, Kyoto 606-8502, Japan
[3] Department of Advanced Materials Science, University of Tokyo, Chiba 277-8561, Japan
[†] Present address: Center for Emergent Matter Science, Riken, Wako, Saitama 351-0198, Japan
[‡] Present address: Research Institute for Interdisciplinary Science, Okayama University, Okayama, 700-8530, Japan
[*] Correspondence: my@issp.u-tokyo.ac.jp

**Abstract:**

A finite residual linear term in the thermal conductivity at zero temperature in insulating magnets indicates the presence of gapless excitations of itinerant quasiparticles, which has been observed in some candidate materials of quantum spin liquids (QSLs). In the organic triangular insulator β'–EtMe$_3$Sb[Pd(dmit)$_2$]$_2$, a QSL candidate material, the low-temperature thermal conductivity depends on the cooling process and the finite residual term is observed only in samples with large thermal conductivity. Moreover, the cooling rate dependence is largely sample dependent. Here we find that, while the low-temperature thermal conductivity significantly depends on the cooling rate, the high-temperature resistivity is almost perfectly independent of the cooling rate. These results indicate that in the samples with the finite residual term, the mean free path of the quasiparticles that carry the heat at low temperatures is governed by disorders, whose characteristic length scale of the distribution is much longer than the electron mean free path that determines the high-temperature resistivity. This explains why recent X-ray diffraction and nuclear magnetic resonance measurements show no cooling rate dependence. Naturally, these measurements are unsuitable for detecting disorders of the length scale relevant for the thermal conductivity, just as they cannot determine the residual resistivity of metals. Present results indicate that very careful experiments are needed when discussing itinerant spin excitations in β'–EtMe$_3$Sb[Pd(dmit)$_2$]$_2$.

## 1. Introduction

Identifying the nature of the elementary excitation plays an essential role in revealing unknown quantum-condensed states. A quantum spin liquid (QSL) [1,2] is a celebrated quantum-condensed state in which the ground state has remained elusive except that a disordered state of the constituent quantum spins persists down to the absolute zero temperature owing to the quantum fluctuations. To reveal the elementary excitations in a QSL, it is indispensable to find the presence or the absence of an energy gap for the excitations. In the former, the presence of the energy gap indicates a formation of a quantum-entangled singlet state with a short correlation length. In the latter, the absence of the gap shows the emergence of itinerant excitations with a long correlation length.

In two dimensions, a QSL is suggested to be stabilized when magnetic order is inhibited by a geometrical frustration of the underlying lattice structure. Indeed, several QSL candidates have been reported in frustrated materials including triangular [3–13] and kagomé antiferromagnets [14–17]. The elementary excitations of these compounds have been studied to find the presence or the absence of an energy gap. In particular, a finite residual of the linear-in-temperature term in the heat capacity [4,8,11] has been paid considerable attention because it is considered as evidence of gapless fermionic excitations. Whereas both localized and itinerant excitations are detected in the heat capacity measurements, only itinerant excitations are probed in thermal conductivity ($\kappa$) measurements. Therefore, a finite residual in the linear-in-temperature term in the thermal conductivity measurement ($\kappa_0/T \equiv \kappa/T$ as $T \to 0$) demonstrates the presence of *itinerant* gapless spin excitations. A finite $\kappa_0/T$ in the QSL candidates has been observed in β'–EtMe$_3$Sb[Pd(dmit)$_2$]$_2$ [7], κ–H$_3$(Cat-EDT-TTF)$_2$ [10], 1T–TaS$_2$ [12], and YbMgGaO$_4$ [13].

Recently, the absence of $\kappa_0/T$ has been reported in some samples of β'–EtMe$_3$Sb[Pd(dmit)$_2$]$_2$ with low thermal conductivity [18–20]. For this discrepancy, it is suggested that suppression effects on the phonon thermal conductivity ($\kappa_{\text{ph}}$) also scatter the spin conduction [18], disabling one to detect $\kappa_0/T$. Indeed, whereas the samples with boundary-limited phonons show a finite $\kappa_0/T$, those with highly-suppressed $\kappa_{\text{ph}}$ do not. Furthermore, the thermal conductivity of some samples is found to depend on the cooling rate [21], implying that extrinsic scatters formed during the cooling process hinder the phonon conduction. However, the absence of the cooling-rate dependence is found in the resistivity, the X-ray diffraction (XRD), and the nuclear magnetic resonance (NMR) measurements [22].

Here, we report that the cooling rate dependence of the low-temperature thermal conductivity is observed despite the resistivity in the measurable temperature range above ~40 K does not show discernible cooling rate dependence. We further find that a finite $\kappa_0/T$ completely disappears by a subsequent rapid cooling performed on the same sample with the same setup. These results demonstrate that the thermal conductivity at low temperatures is sensitive to disorders formed in a macroscopic length scale that does not affect the high-temperature resistivity, the XRD, and the NMR measurements.

## 2. Results
### *Cooling-rate dependence of the resistivity*
The temperature dependence of the resistivity of Sample #2–#4 is shown in Fig. 1. The data taken from Ref. [22] is also shown. The data is normalized by the value at 300 K to avoid the errors brought by the sample dependence and the estimation of the sample dimensions. The cooling rate of Sample #2, #3, and #4 was −1.5, −13, and −85 K/h, respectively. The cooling rate of the data of Ref. [22] is −0.6 to −150 K/h. As shown in Fig. 1, the temperature dependence of the normalized resistivity of all the samples is virtually overlapped, showing the absence of the cooling-rate dependence of the resistivity as reported in Ref. [22]. Note that below ~40 K, the resistivity becomes too large for accurate measurements, owing to the insulating nature of this material. As shown by the Arrhenius plot of the data (the inset of Fig. 1), the temperature dependence of the resistivity follows the activation behavior ($\propto \exp(-\Delta/k_B T)$) with the energy gap of $\Delta \sim 240$ K, showing that the electric transport is well described by a simple Mott insulating model.

### *Cooling-rate dependence of the thermal conductivity*
Figure 2 shows the temperature dependence of the thermal conductivity divided by the temperature ($\kappa/T$) of Sample #2–#4 below 2.5 K. As shown in Fig. 2, in spite of almost the same temperature dependence of the resistivity, the thermal conductivity of these samples differs each other about a factor of 2–3. Moreover, whereas a linear extrapolation of $\kappa/T$ to $T = 0$ (dashed lines in Fig. 2) shows a finite residual $\kappa_0/T$ for Sample #2, those of Sample #3 and #4 are vanishingly small, showing a cooling rate effect on $\kappa_0/T$.

This contrasting difference in the cooling-rate dependence of the high-temperature resistivity and that of the low-temperature thermal conductivity demonstrates that neither the magnitude of the thermal conductivity nor the presence and the absence of $\kappa_0/T$ are related to the temperature dependence of resistivity.

***Subsequent rapid cooling effect on the thermal conductivity of the same sample***
The temperature dependence of $\kappa/T$ of Sample #1 is shown in Fig. 3. This sample was cooled down by the slowest cooling rate ($-0.4$ K/h) in the initial cooling run. As shown in Fig. 3, the magnitude of $\kappa/T$ measured after the slow cooling rate (red diamonds) is the largest in these four samples. In addition, the residual $\kappa_0/T$ estimated by the linear extrapolation (the red dashed line in Fig. 3) also shows the largest value.

To check the effect of the cooling rate on the same sample, the subsequent thermal cycle with a rapid cooling rate of $-150$ K/h was applied on Sample #1 after warming up the sample to room temperature without changing any setups. Remarkably, as shown in Fig. 3, $\kappa/T$ of Sample #1 is strikingly suppressed by the rapid cooling and the finite $\kappa_0/T$ disappears (blue diamonds in Fig. 3), demonstrating the cooling rate effect on the magnitude of the thermal conductivity and that on $\kappa_0/T$. Although the detailed temperature dependence of the resistivity of Sample 1 was not measured at low temperatures due to the setup that was not optimized for high-resistivity measurements, we confirmed that the resistivity was not changed at room temperature by these thermal cycles. This is also consistent with the result that the scatters for the thermal transport do not affect the electric transport.

**Discussion**
Our results indicate that the temperature dependence of the resistivity of β'–EtMe$_3$Sb[Pd(dmit)$_2$]$_2$ at high temperatures is irrelevant to the thermal-transport properties at low temperatures of the same sample. This can be understood by considering that different carriers are responsible for the electrical and thermal transport in this QSL system.

Electric transport is carried by electrons. As shown by the Arrhenius plot of the resistivity data (the inset of Fig. 1), this organic material is a Mott insulator with an electron in each Pd(dmit)$_2$ dimer. Its electric transport is thus completely dominated by activated hopping processes between the neighboring dimers. In fact, by estimating the carrier density at the room-temperature as one electron per dimer ($\sim 2 \times 10^{18}$ m$^{-2}$ from the lattice constants of the $a$–$b$ plane[23]), one can obtain an extremely short mean free path of electron (order of $\sim 10^{-11}$ m) from the room-temperature resistivity of $\rho \sim 0.1$ $\Omega \cdot$cm. Therefore, the absence of the cooling rate dependence in the resistivity [22] shows the absence of the change in the length scale of molecules during the cooling, except those caused by the lattice contraction. The cooling-rate dependence investigated by the XRD and NMR measurements [22] essentially shows the same absence of the change in the length scale of the molecule.

In sharp contrast, thermal transport of an insulator is mainly carried by phonons. Phonons are bosons of which the wavelength becomes longer as lowering the temperature. Thus, the length scale related to the phonon conduction becomes a macroscopic length (> μm) at low temperatures. As a result, phonon conductions are only scattered by macroscopic objects such as boundaries at low temperatures. The mean free path of phonons ($\ell_{\rm ph}$) of this compound is determined by $\kappa_{\rm ph}$ as $\kappa_{\rm ph} = C_{\rm ph} v_{\rm ph} \ell_{\rm ph}/3$, where $C_{\rm ph}$ and $v_{\rm ph}$ are the heat capacity and the velocity of the phonons, respectively. Since no magnon contribution is expected in this QSL compound[6], the thermal conductivity is given by the sum of $\kappa_{\rm ph}$ and the itinerant spin excitations given by $\kappa_0/T$. We evaluate $\kappa_{\rm ph}$ by subtracting the itinerant spin contribution from the measured $\kappa$. We then estimate $\ell_{\rm ph}$ by using $C_{\rm ph}/T^3 = 22.8$ mJ K$^{-4}$ mol$^{-1}$ from the specific heat data[8] and $v_{\rm ph} = 2,100$ m/s[18]. As shown in Fig. 4, $\ell_{\rm ph}$ of the measured samples indeed reaches 1–30 μm at the lowest temperature [18], demonstrating the macroscopic length scale relevant to the thermal transport. Therefore, the different magnitude of the thermal conductivity caused by the different cooling rates shows the formations of random scatters with different macroscopic length scales during the cooling process. These scatters for phonons do not affect the electric transport because the electric transport is already dominated by the activated hopping processes between the neighboring dimers. The XRD and NMR measurements are also not affected because these probes detect the spatially averaged property of the constituent atoms. As a result, all the cooling rate dependence reported in Ref. [22] is irrelevant to the thermal transport at low temperatures.

Having established the irrelevance between the high-temperature resistivity and the low-temperature thermal conductivity, we now discuss that the presence and the absence of $\kappa_0/T$ strongly depends on the magnitude of the phonon thermal conductivity. This is most clearly shown in Fig. 3 in which the magnitude of the thermal conductivity is strongly suppressed in the subsequent rapid cooling despite the same sample setups. At the same time, the finite $\kappa_0/T$ observed in the first slow-cooled run completely disappears by the subsequent rapid cooling. This presence and the absence of $\kappa_0/T$ measured in the same sample demonstrate that the presence of the gapless excitations is the intrinsic property of the QSL state in this material and that the absence of $\kappa_0/T$ is caused by extrinsic scatters formed by the rapid cooling. Similar sample dependence of $\kappa_0/T$ has also been reported in 1T-TaS$_2$ [12] and YbMgGaO$_4$ [13], in which the presence (absence) of $\kappa_0/T$ is observed in a high (low) $\kappa$ sample, showing that the gapless spin excitations are sensitive to the scatters of phonons. These results demonstrate that the presence and the absence of the gapless excitations must be studied in samples with sufficiently high phonon conduction. Otherwise, it is indistinguishable if the gapless excitations do not exist or they are simply suppressed by extrinsic scatters in the sample. We note that the formation of extrinsic scatters during the cooling shows a sample variation. As shown in Ref. [21], a finite $\kappa_0/T$ is observed in samples cooled by a cooling rate comparable to that used in sample #3. The magnitude of $\kappa_0/T$ also shows a large sample variation even in samples cooled down a similar cooling rate under a similar setup [7,18,21]. This complicated sample variation shows that the formation of the extrinsic scatters is not determined only by the cooling rate. The other factors, such as stress applied on the sample by the setup for the thermal-transport measurements and the large thermal contraction of the organic compound [24], would also contribute to the formation of the extrinsic scatters.

**Conclusions**

From the cooling-rate dependence of the resistivity and the thermal conductivity, we show that the low-temperature thermal conductivity depends on the cooling rate despite the independence of the resistivity. This difference is brought by the longer length scale determining the low-temperature thermal conductivity than that for the resistivity, which also naturally explains the absence of the cooling rate dependence of XRD and NMR measurements. We further show that, even in the same sample under the same setups, the rapid cooling drastically suppresses the thermal conductivity, obscuring the detection of the residual of $\kappa/T$ in the zero-temperature limit. Moreover, the cooling rate-dependence of the low-temperature thermal conductivity also depends on the samples. These results indicate that very careful experiments are needed when discussing itinerant spin excitations in β'–EtMe$_3$Sb[Pd(dmit)$_2$]$_2$.

**Figures and Legends:**

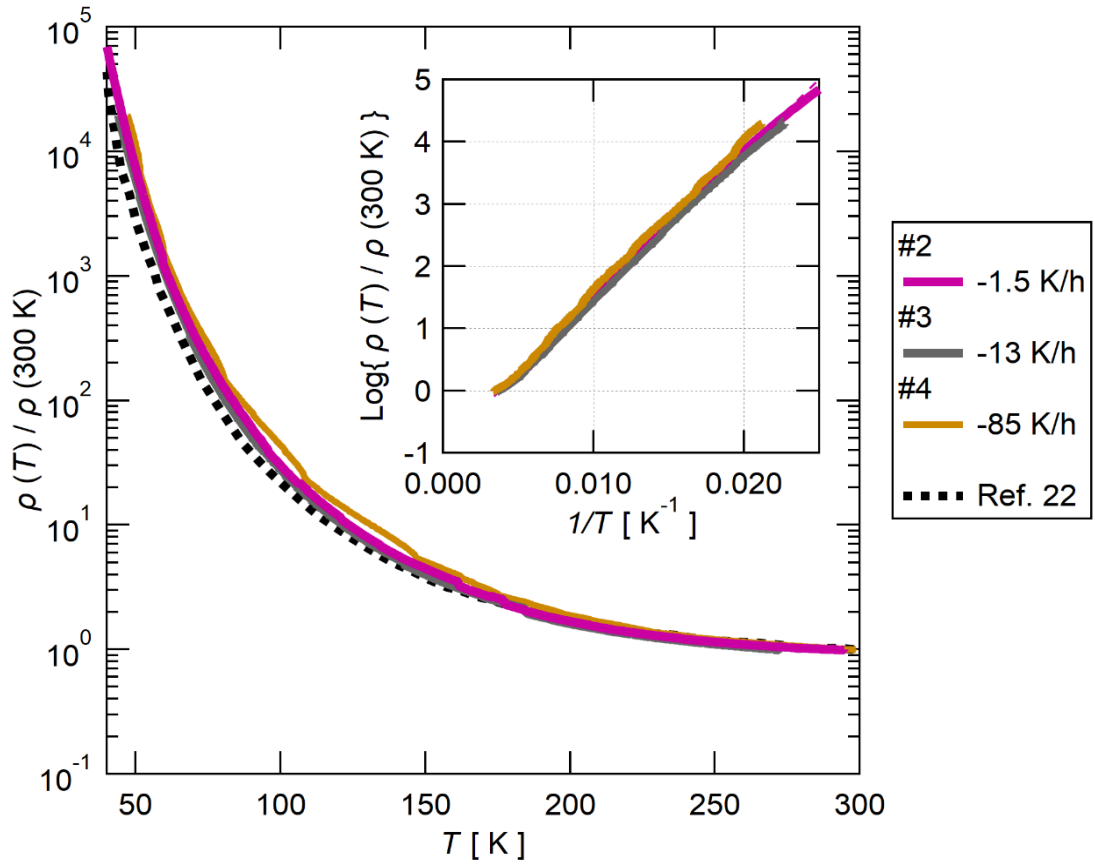

**Figure 1.** The temperature dependence of the resistivity of Sample #2, #3, and #4 normalized by the value at room temperature. The data of Ref. [22] is shown for reference (the dashed line). The inset shows the Arrhenius plot of the data, showing that the temperature dependence of the resistivity follows the simple activated behavior ($\rho \propto \exp(-\Delta/k_\mathrm{B}T)$).

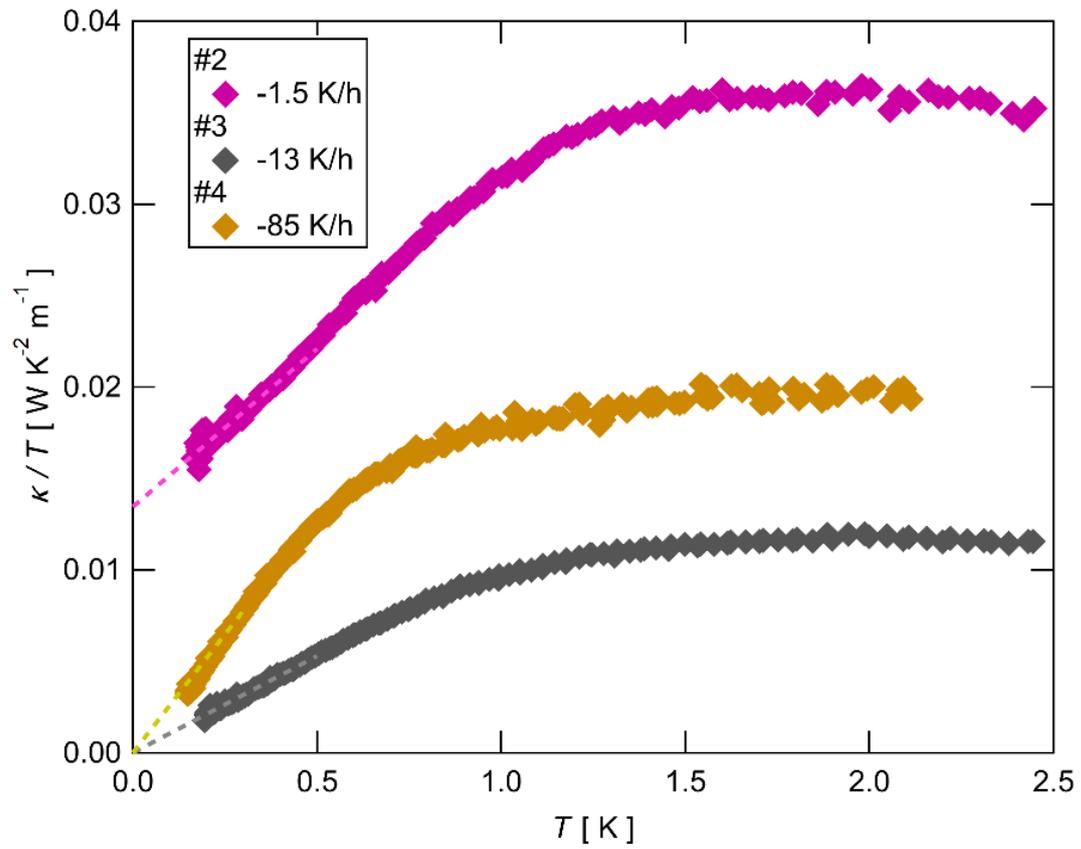

**Figure 2.** The temperature dependence of the thermal conductivity divided by the temperature ($\kappa/T$) of Sample #2, #3 and #4. The data of Sample #2 and #3 is the same data in Ref. [21]. The dashed lines show a linear fit of the data at lower temperatures.

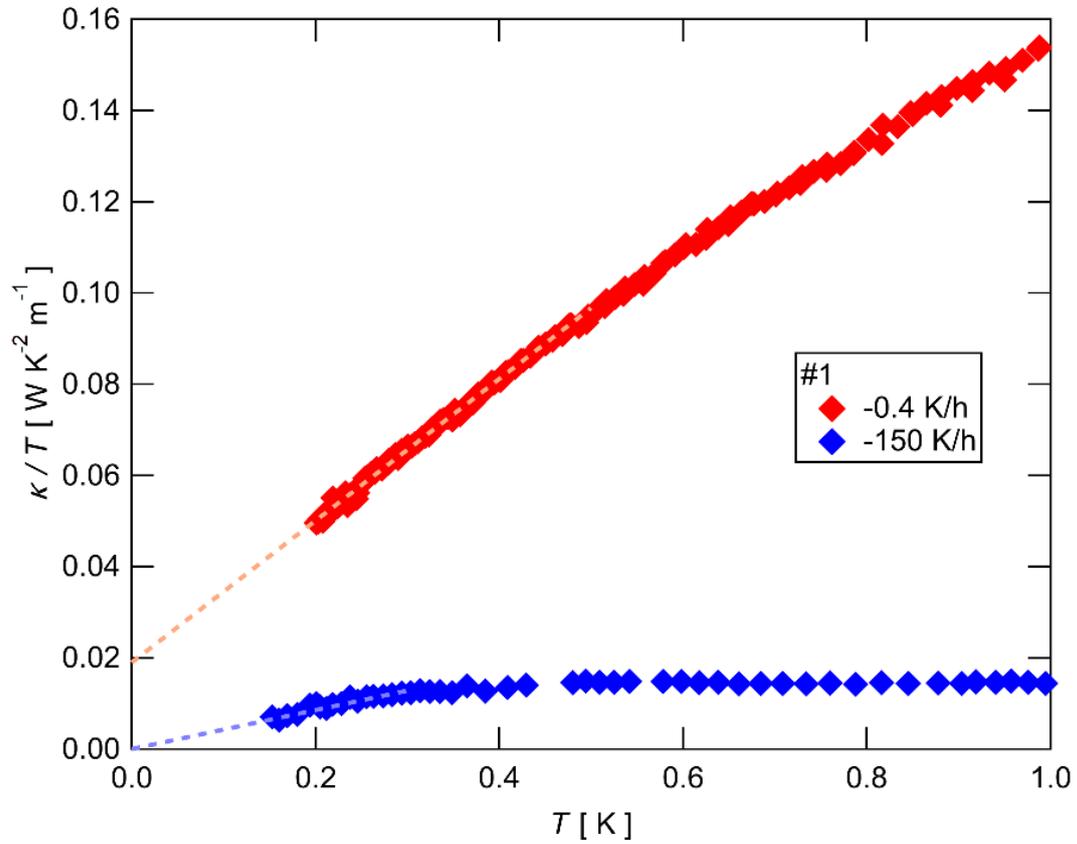

**Figure 3.** The temperature dependence of the thermal conductivity divided by the temperature ($\kappa/T$) of Sample #1. The data of the first (-0.4 K/h) and the second (-150 K/h) cooling process is shown by the red and the blue symbols, respectively. The data of the first cooling of Sample #1 is the same data in Ref. [21]. The dashed lines show a linear extrapolation of the data.

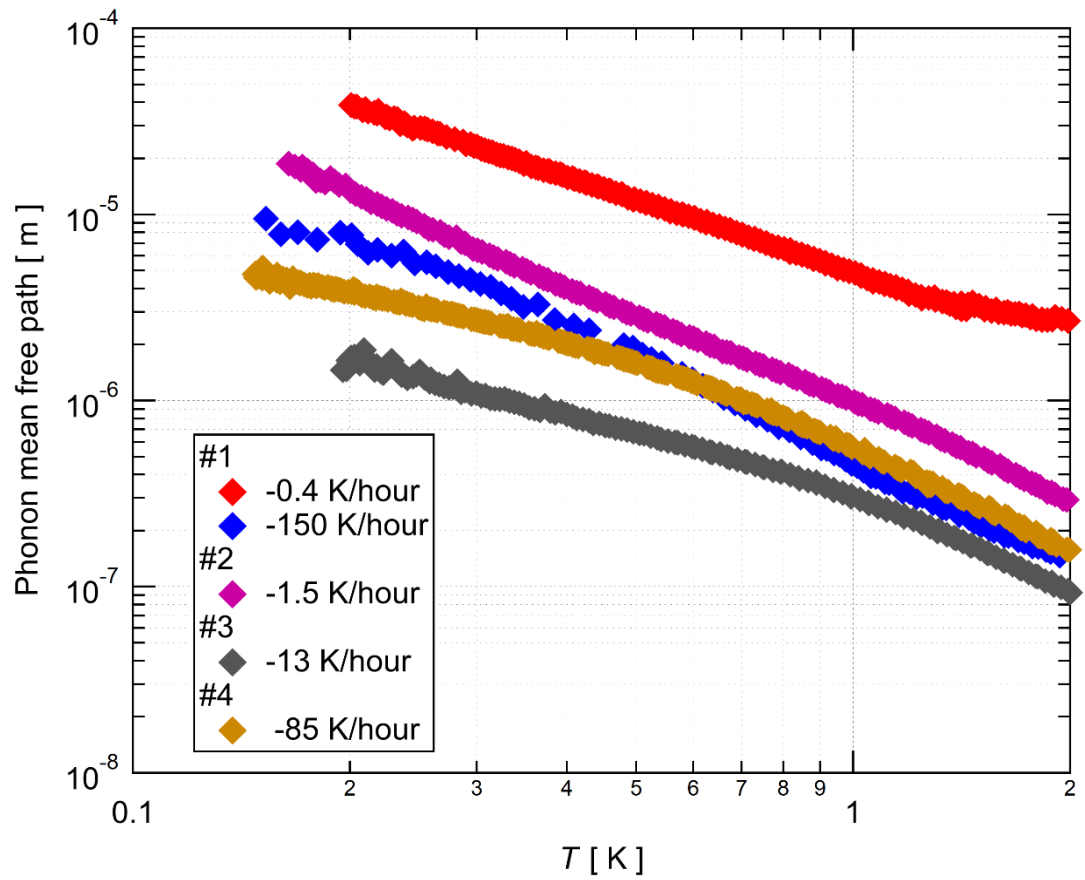

**Figure 4.** The temperature dependence of the phonon mean free path. See the main text for details.

**Materials and Methods:**

The samples were synthesized by R. Kato as described in Ref. [22]. The electrical resistivity and the thermal conductivity were measured in the same samples during the same cooling runs.

The resistivity was measured by the standard four-probe method using a digital multimeter with a dc current (10–100 μA) applied within the $a$–$b$ plane of the sample. The thermal conductivity was measured by the steady-state method using one chip heater and two calibrated $RuO_2$ thermometers. These heaters and thermometers were thermally connected to the sample by attaching gold wires by a carbon paste. These thermal contacts were also used as the electric contacts for the resistivity measurements. The heat current, applied to the $a$–$b$ plane of the sample, was adjusted to make the longitudinal temperature difference 1–3% of the sample temperature.

The samples (#1–4) were cooled down at a steady cooling rate of $-0.4$, $-1.5$, $-13$, and $-85$ K/h from room temperature, respectively. The temperature dependence of the resistivity was recorded during the cooling. After reaching the lowest temperature, the temperature dependence of the thermal conductivity was measured. Sample #1 was cooled again by a rapid cooling rate of $-150$ K/h after warming up to room temperature.

The standard error of all the resistivity and the thermal conductivity data is smaller than the symbol size of the figures. The uncertainty in the absolute value of the thermal conductivity data is caused by the ambiguity in estimating the distance between the thermal contacts and the thickness of the sample. This uncertainty is estimated as approximately 10%.

**Data Availability:**
All the data that support the findings of this study are available from the corresponding author (M.Y.) upon reasonable request.

**Acknowledgements:**
We acknowledge R. Kato for providing single crystals used in this work. This research was supported by Grants-in-Aid for Scientific Research (KAKENHI) (No. JP19K21842 and JP19H01848) and on innovative areas "Quantum Liquid Crystals" (No. JP19H05824) from the Japan Society for the Promotion of Science.